\documentclass[12pt,a4paper]{article}

\usepackage{amsmath,amssymb}
\usepackage{epsfig}
\newcommand{\bu}{$\bullet$\ }
\newcommand{\be}{\begin{equation}}
\newcommand{\ee}{\end{equation}}
\newcommand{\bes}{\begin{subequations}}
\newcommand{\ees}{\end{subequations}}
\newcommand{\bea}{\begin{eqnarray}}
\newcommand{\eea}{\end{eqnarray}}
\newcommand{\bear}{\begin{equation}\begin{array}}
\newcommand{\eear}[1]{\end{array}\label{#1}\end{equation}}
\usepackage{wrapfig}

\newcommand{\ls}[1]{\lambda_{#1}}
\def\ba{$$\begin{array}}
 \def\ea{\end{array}$$}

\newcommand{\fr}[2]{\dfrac{{ #1}}{{ #2}}}
\newcommand{\pa}{\partial}
\newcommand{\la}{\langle}
\newcommand{\ra}{\rangle}
\newcommand{\fn}[1]{\footnote{{#1}}}

\def\vep{{\varepsilon}}

{\end{list}}
\newcounter{enumct}

\newcommand{\LL}[1]{\Lambda_{#1}}

\newcommand{\ff}[2]{(\!\phi_#1^\dagger\phi_#2\!)}

\newcommand{\Ls}[1]{\Lambda_{#1}}

\begin{document}
\title{\bf Nonminimal Higgs Models, Dark Matter,
and Evolution of the Universe$^a$
}

\author{I.~F.~Ginzburg
\\
{\small Sobolev Institute of Mathematics, Novosibirsk, 630090, Russia;}\\{\small Novosibirsk State University, Novosibirsk, 630090, Russia;}\\{\small\it  E-mail: ginzburg@math.nsc.ru}}
\date{}

\maketitle

\begin{abstract}
The set of sum rules for a wide class of nonminimal Higgs models has been obtained. Difficulties and ways for
revealing the possibilities of studying extended Higgs models at colliders have been revealed with the use of
these sum rules and recent LHC results. New methods of studying multidoublet Higgs models with various
symmetry groups have been applied to solve problems of classification of these groups, breaking of symmetries in vacuum, etc. A method for the determination of masses and spins of dark matter particles D and their
partners via the energy spectrum of a lepton in the $e^+e^-\to DDW^+W^-$ process has been proposed. The possibility of the existence of strongly interacting dark matter has been revealed. Variants of the evolution of the
phase states of the Universe have been analyzed within the inert doublet model.

$^a$ {\it With correction some misprints in journal publication JETP Letters {\bf  99}  (2014) 742-751.}
\end{abstract}

\section{Introduction}
Nonminimal models of electroweak symmetry
breaking -- models with additional Higgs bosons -- are
discussed in this work with the focus on difficulties and
possibilities of the experimental observation of new
particles predicted in such theories taking into
account recent data on the properties of the Higgs
boson obtained at the LHC (Sections 2-5).
A special variant of a Two Higgs Doublet Model--
the inert doublet model (IDM)-naturally introduces
candidates for dark matter particles. A method for the
measurement of the masses of these particles at the
ILC linear collider has been discussed. It is found that
strongly interacting dark matter can exist within this
model (Sections 6 and 7).
The parameters of the potential of the IDM vary in
the process of cooling of the Universe in the postinflation era and the phase states of the Universe can
change so that, after the usually discussed phase transition breaking electroweak symmetry, the Universe
can undergo an additional first order phase transition
or a pair of second order phase transitions at fairly low
temperatures with a rich structure of critical fluctuations (Sections 8 and 9).

\section{Higgs Models}\label{secintro}

The electroweak theory describes experimental
data well. To complete the development of the theory,
it remained to reveal how electroweak symmetry
breaking (EWSB) occurs. After the detection of the
Higgs boson with the mass $M_h=125$\,GeV \cite{125Higgs}, the
physics community believes that EWSB is due to the
Higgs mechanism with the Lagrangian
\be
{ \cal L}={ \cal L}^{SM}_{ gf }\! +\! {\cal L}_Y \!+T -V.\label{baslagr}
 \ee
Here ${ \cal L}^{SM}_{ gf }$ describes the standard $SU(2)\times U(1)$ interaction of the gauge bosons and fermions, ${\cal L}_Y$  is the
Yukawa interaction of fermions with scalar Higgs
fields  $\phi_i$, and the last two terms describe the Higgs
fields ($T=\sum_i ({\cal D}_\mu\phi_i) ({\cal D}_\mu\phi_i)^\dagger/2$ is the standard
kinetic term and $V$ is the Higgs potential).

The minimal variant (Standard Model) includes
one fundamental Higgs field $\phi$ (weak isodoublet) and
$V=-m^2(\phi^\dagger\phi)/2+\lambda(\phi^\dagger\phi)^2/2$. The field $\phi$ has four
degrees of freedom. After EWSB, three components of
this field remain massless. These are Goldstone
modes, which become the longitudinal components of
gauge fields. One component is manifested as a scalar
Higgs boson whose mass is an almost arbitrary parameter of the theory. The coupling constants of the Higgs
boson with fundamental particles are unambiguously
determined.

The same EWSB mechanism can also be implemented in nonminimal models containing several fundamental Higgs fields $\phi_i$. They allows the natural
description of the CP symmetry breaking and flavor
changing neutral currents, as well as the introduction
of candidates for dark matter particles, etc. (Such
models naturally appear in, e.g., models of broken
supersymmetry for the description of reality: MSSM,
nMSSM, etc.).  Models nHDM + pHSnM containing n
scalar isodoublets and p scalar isosinglets will be specially considered below. (The Higgs sectors in MSSM
and nMSSM have the form 2HDM and 2HDM +
1HSnM, respectively.)

\bu {\textbf The relative coupling constants} of gauge bosons $V =
Wm\, Z$, leptons and quarks $f$ with the neutral Higgs
boson $h_a$ are the ratios of the observed constant values
to the respective values in the Standard Model (SM):
\bes\label{relcoupl}\be
\chi_V^{(a)}=g_V^{(a)}/g_V^{\rm SM}, \quad
\chi_f^{(a)}=g_f^{(a)}/g_f^{\rm SM}\,.\label{relcoupl1}
\ee

The nonminimal Higgs models include additional
interactions of scalar and vector bosons. In models
with one charged Higgs boson $H^\pm$, the quantities
\begin{equation}
    \chi^{(a)}_{H^\pm W^\mp} = \dfrac{g(H^\pm W^\mp h_a)}{M_W/v},\quad \chi^{(ab)}_{Z} = \dfrac{g(Z h_a h_b)}{M_Z/v}.\label{relcoupb}
\end{equation}
\ees
are introduced.

\bu {\bf SM-like situation.} After EWSB, new fundamental
fields appear in the form of a set of charged and neutral
scalar Higgs bosons H± and ha. The observation of
these new bosons will be the main indication of the
realization of nonminimal models. The deviation of
the coupling constant values from their values in the
Standard Model will be a preliminary signal of the
realization of such a model.

In view of this circumstance, the following question arises. {\it Let the experimental situation at a certain
time be indistinguishable within the experimental accuracy from that predicted by the Standard Model; i.e.,
only one Higgs boson is detected and its interactions with
fundamental particles within the experimental accuracy
$\vep^{exp}$ do not differ from the predictions of the Standard
Model}:
\be
||\chi_V^{(1)}|-1|,\quad||\chi_f^{(1)}|-1|<\vep^{exp}_{V,f}\,.\label{SMLsc}
\ee

{\it This situation is called by us the {\bf SM-like situation}
\cite{GKO}. Is this compatible with the realization of some nonminimal theory? What are the "simplest" experiments
that make it possible to distinguish these possibilities?}

The SM-like situation in the nonminimal model
can occur if additional Higgs bosons are very heavy
and are coupled only weakly with usual matter (decoupling limit). In \cite{GKO} it was found that, at the experimental accuracy expected at the LHC, which was being
built at that time, and at the planned linear $e^+e^-$ collider, even the simplest nonminimal model 2HDM
with the special choice of the Yukawa interaction
2HDM-II (as in MSSM) allows several possibility
windows significantly differing from the decoupling
limit and implementing the SM-like situation. It is
clear that such windows exist in other models as well.
(At the same time, study of the production of the
Higgs boson at a photon collider will promote the
choice between the models \cite{GKO}.)

Successful experiments (decrease in $\vep^{exp}$) reduce
the region of the allowed parameters of nonminimal
model \eqref{SMLsc}.

The LHC experiments indicate the realization of
the SM-like situation \cite{125Higgs}, \cite{125_2HDM} (with inaccuracies much
larger than those discussed in \cite{GKO}). The world physics
community is now actively seeking such possibilities
and probable signals of deviations from the Standard
Model in future experiments.

\section{Two Higgs doublet model}\label{sec2HDM}

The simplest alternative to the minimal variant of
the Standard Model is the model with two Higgs doublets $\phi_1$ and $\phi_2$ (2HDM). Its potential is generally
described by 14 parameters (4 complex parameters $m_{12}^2$
and $\lambda_{5-7}$ and real valued remaining parameters):
\bear{c}
V\!=\!-\dfrac{1}{2}\left[m_{11}^2(\phi_1^\dagger\phi_1)\!+
	m_{22}^2(\phi_2^\dagger\phi_2)\right.
\left.\!+\!\left( m_{12}^2(\phi_1^\dagger\phi_2)\!+\!h.c.\right)
\right]\\[2mm]
+\dfrac{\lambda_1(\phi_1^\dagger\phi_1)^2 \!+\!\lambda_2(\phi_2^\dagger\phi_2)^2}{2}+ \lambda_3(\phi_1^\dagger\phi_1)(\phi_2^\dagger\phi_2)
	\!+\!\lambda_4(\phi_1^\dagger\phi_2)(\phi_2^\dagger\phi_1)\\[2mm] +\left(\dfrac{\lambda_5}{2}(\phi_1^\dagger\phi_2)^2!+
\!\lambda_6(\phi_1^\dagger\phi_2)(\phi_1^\dagger\phi_1)
\!+\! \lambda_7(\phi_1^\dagger\phi_2)(\phi_2^\dagger\phi_2)+h.c.\right).
\eear{baspot}
Different variants of the model differ in the form of the
Yukawa interaction ${\cal L}_Y$.

Owing to the presence of two fields with identical
quantum numbers in the model, the same physical
reality can be described by different forms of the
Lagrangian. They can be transformed to each other by
a global linear transformation of the fundamental
Higgs fields (generalized rotation$\phi_1, \phi_2\to \phi_1^\prime, \phi_2^\prime$)
with the appropriate variation of the parameters of the
potential {\it(reparameterization (RPa) invariance)}. The
mentioned rotation is described by three (gauge)
parameters. Consequently, it is sufficient to determine
11 significant parameters for the complete description
of the model.

The potential has a minimum at certain values $\la\phi_i\ra$
of classical fields. The isotopic direction corresponding to the neutral field along the lower component of
the weak isospinor $\la\phi_i\ra$
 is usually chosen. Correspondingly, the minimum conserving the charge has the
form
\be
\la\phi_1\ra=\dfrac{1}{\sqrt{2}}\begin{pmatrix}0\\v_1\end{pmatrix}\,,\quad
\la\phi_2\ra=\dfrac{1}{\sqrt{2}}\begin{pmatrix}0\\v_2e^{i\xi}\end{pmatrix}\,. \label{vev1}
\ee

Under RPa rotation, the ratio $v_2/v-1 = \tan\beta$ varies
and $v =\sqrt{v_1^2+v_2^2} = 246$~GeV is conserved.

Different RPa gauges are convenient for solving
different physical problems. In particular, for studying
consequences of some symmetry, it is preferable to use
a gauge in which this symmetry is written in the simplest form.

The Higgs basis in which
$\la\phi_1\ra =\dfrac{1}{\sqrt{2}}\begin{pmatrix}0\\v\end{pmatrix}$, and
$\la\phi_2\ra=0$
 is convenient in a number of problems (see,
e.g. \cite{Nishi}). This basis is obtained from the basis in which
v.e.v.'s have the form of Eqs. \eqref{vev1} by the
transformation
\be
\begin{pmatrix}\phi_{1H}\\\phi_{2H}\end{pmatrix}\!\!=\!\!
\begin{pmatrix}
     \cos\beta\,e^{i\rho/2}       &\!\!\sin\beta\,e^{i(\rho/2-\xi)}\!\!\!\\
    \!\!-\sin\beta\,e^{-i(\rho/2-\xi)}\!\!&\cos\beta\,e^{-i\rho/2}\!
    \end{pmatrix}\!\!
\begin{pmatrix}\phi_1\\\phi_2\end{pmatrix}\label{RPaHB}
\ee
with the appropriate change in the parameters of the
potential $\lambda_i\to \Lambda_i$ (below, the subscript $H$ will be
omitted).

The potential can be represented in the form
including the mass of the charged Higgs boson $M_\pm$ :
\bear{c}
V_{HB} = M_\pm^2\ff22\! +\! \dfrac{\Ls1}{2}\left(\ff 11-\fr{v^2}{2}\right)^2
\!+\!\dfrac{\Ls2}{2}\ff22^2
+ \Ls3\left(\ff 11-\fr{v^2}{2}\right)\ff22\\[2mm]+\Ls4\ff12\ff21
        + \left[\dfrac{\Ls5}{2}\ff12^2+\Ls6\left(\ff 11-\fr{v^2}{2}\right)\ff12 
 + \Ls7\ff22\ff12+\text{h.c.}\right].
\eear{HBmpot2}
This potential holds its form under the {\it rephasing
(RPh) transformation} $\ff12\to \ff12 e^{i\rho}$ \eqref{RPaHB}
with
the appropriate change in the parameters $\LL i$.

Then, the fields are expanded in terms of deviations
from the vacuum average:

\begin{equation}
\phi_1=\left(\begin{array}{c}G^+\\ \dfrac{v+\eta_1+iG^0}{\sqrt2}\end{array}\right),
\;\;
\phi_2=\left(\begin{array}{c}H^+\\ \dfrac{\eta_2+i\eta_3}{\sqrt2}\end{array}\right).
\label{decomp}
\end{equation}
Here, $G^\pm$ and $G^0$ are the Goldstone fields transformed
to the longitudinal components of massive gauge fields
$W^\pm$ and $Z$ (they are omitted below), $H^\pm$ are the
charged Higgs fields with the mass $M_\pm$, and neutral
Higgs fields $h_a$ are formed from neutral components $\eta_i$:
\be
h_a=R_a^i\eta_i\,.\label{etatoh}
\ee

The elements of the mixing matrix $R_a^i$ are real valued.
A similar relation can certainly be written in any RPa
basis. The advantage (for analysis) of the Higgs basis is
that the measurable physical quantities, coupling constants \eqref{relcoupl}  are directly expressed in terms of $R_a^i$:
\bear{c}
    \chi^{(a)}_V =R_a^1, \quad \chi^{(a)}_{H^\pm W^\mp} = R_a^2+iR_a^3,\\[2mm]
    \chi^{ab}_Z = R_a^2R_b^3-R_b^2R_a^3.
    \eear{gcoupl}
(The phases of the quantities $\chi^{(a)}_{H^\pm W^\mp}$, i.e., the ratios $R_a^3/R_a^2$, cannot be fixed because of the presence of
phase freedom in the definition of fields in Eq.~\eqref{HBmpot2}, but
their relative phases are unambiguously defined. In
particular, the gauge of this phase can be fixed by the
requirement that one of the parameters $\chi^{(a)}_{H^\pm W^\mp}$ be real valued.)

The direct consideration of Eq.~\eqref{HBmpot2} makes it possible to verify that the parameters of the model are classified into two groups. The parameters in the first
group ($\LL1,\, \LL4,\, \LL5,, \LL6$, $M^2_\pm$, $v^2$) are completely determined
by the masses of scalar particles and their coupling
constants with gauge bosons. The remaining parameters ($\LL3,\, \LL7,\, \LL2$) cannot be determined without the
measurement of triple and quadruple interactions of
Higgs bosons (the $H^+H^-h_a$ and $H^+H^-H^+H^-$ vertices
are the best candidates for these interactions). Furthermore, even at moderate masses of Higgs bosons, a
scenario with large $\LL2$ values is possible (strong interaction in the Higgs sector, which should be studied
separately   \cite{53}). This is the difference from the Standard Model, where the physical Higgs boson disappears under the strong coupling conditions and the
width of the Higgs boson becomes comparable to its
mass.

The orthogonality of the mixing matrix means that
$
\sum\limits_i |R^a_i|^2=1$ and $\sum\limits_a |R^a_i|^2=1$. These relations
together with \eqref{gcoupl} can be represented in the form
of the sum rules \cite{53}
\bear{c}
a)\; \sum\limits_i|\chi^{(a)}_V|^2=1\,,\\[2mm]
b)\; |\chi^{(a)}_V|^2+| \chi^{(a)}_{H^\pm W^\mp}|^2=1,\;\; \sum\limits_i|\chi^{(a)}_{H^\pm W^\mp}|^2=2,\\[2mm]
c)\; \chi^{(ab)}_Z=Im\left(\chi^{*(a)}_{H^\pm W^\mp}\chi^{(b)}_{H^\pm W^\mp}\right)
 \,.
\eear{SR}

\section{Sum rules and possibilities
of future experiments}

The presented method of obtaining sum rules
makes it possible to expand these rules to much wider
classes of models.

\bu Sum rule (\ref{SR}a) is well known in 2HDM \cite{gunion-haber-wudka,GKO,GK05}
and means that the masses of gauge bosons are determined by the Higgs mechanism of EWSB. For this reason, sum rule  (\ref{SR}a) is valid in any nonminimal Higgs
model both such that $\chi_W^{(a)}= \chi_Z^{(a)}$  (as in the Standard
Model and nHDM) and such that $\chi_W^{(a)}\neq\chi_Z^{(a)}$  (e.g.,
models with additional isosinglet and (or) isotriplet
Higgs fields) \cite{GinKr13}.

Sum rules (\ref{SR}b) and relation (\ref{SR}c) are formulated
for the first time. They are valid for all models whose
physical sector includes only one charged Higgs boson
($H^\pm$). These are models including two isodoublets and
p isosinglets, 2HDM + pHSnM, in particular,
nMSSM.

Sum rules (\ref{SR}a)-(\ref{SR}c) describe the Higgs sector of
theory at any type of (Yukawa) interaction with fermions. Sum rules are known for ({\it generally complex})
coupling constants of a certain fermion (quark or lepton) $f$ with neutral Higgs bosons $\chi_f^{(a)}$ in models with
two Higgs doublets and a certain type of their interaction with fermions, 2HDM-II and 2HDM-I \cite{gunion-haber-wudka,GKO,GK05}:
 \be
\sum\limits_{a}(\chi_f^{(a)})^2=1\,.
\label{vsr}
 \ee

\bu We found that these sum rules are extended for the
nHDM + pHSnM models with arbitrary n and p values
if weak isoscalar fields are not coupled to fermions and
without additional limitations for the form of the
Yukawa interaction \cite{GinKr13}. To prove this statement, let us
write the general interaction of the fermion $f$ with the
fundamental Higgs isodoublet in the form $\Delta L_Y =\sum_j g_{jf}\bar{\psi}^\dagger\phi_j \psi_f$.
The simple reparameterization $\phi_1^\prime=N\sum_j g_{jf}\phi_j$
(where $N$ is the normalization factor)
transforms this contribution to the form $\Delta L_Y = g_{1f}^\prime\bar{\psi}^\dagger\phi_1^\prime \psi_f$, which coincides with the form of
the corresponding interaction in 2HDM-II (or
2HDM-I), where sum rule \eqref{vsr} has already been
proven.

Relations between Yukawa constants for different
fermions appear only under special assumptions on
the structure of this interaction.

\bu {\bf On the strategy of search for new Higgs bosons at colliders in the SM-like situation.}

The main problem of
the verification of nonminimal models is search for
new neutral Higgs bosons $h_a$ with masses $M_a$ and
widths $\Gamma_a$ (we consider new Higgs bosons different
from the already detected one, $a \neq 1$, and having
masses $M_a > 150$~GeV). In various particular models,
such problems have been discussed for many years, as
a rule, for certain masses of these particles. Until
recently, it was commonly expected that the properties
of this boson $h_a$ are close to those of the would be standard Higgs
boson in the Standard Model with approximately the
same mass. In view of the SM-like situation, these
expectations are unjustified in any nonminimal
model.

{\it Decay channels and widths $\Gamma_a$ a of Higgs bosons $h_a$}. In the
Standard Model, the main contribution to the width
of the Higgs boson with masses larger than 150 GeV
comes from the decays $h\to W^+W^-$ and $h\to ZZ$.
These decays would give the main signal of the detection of the CP even Higgs boson. According to sum
rule (\ref{SR}a), the coupling constants of new Higgs bosons
$h_a$ are small. Therefore, the detection of bosons $h_a$
through their decays into gauge bosons is highly
improbable (while detection through decays into
quarks is difficult because of a large background). For
the same reason, the widths $\Gamma_a$  at $Ma < 350$~GeV are
very small {\it in any nonminimal Higgs model}. In this case,
the main decay channel is again $h_a\to \bar{b}b$ with very
large background; consequently, the detection of each
$h_a$ is a difficult problem.

At $Ma > 350$~GeV, the $h_a\to \bar{t}t$ decay channel is
open. Sum rule \eqref{vsr} includes the coupling constants
$\chi_t^{(a)}$, which are generally complex. The smallness of
the sum $\sum\limits_{a>1}(\chi_t^{(a)})^2=1-(\chi_t^{(1)})^2$ can be ensured at
large individual terms (e.g., one is real and the other
is imaginary). Such an example occurs in 2HDM-II
at $\tan\beta < 1$ (values $\tan\beta < 1/7$ should be excluded
because this is the region of strong Yukawa interaction, where perturbative estimates are invalid). In this
case, the ratios $\Gamma_a/M_a$ at least for a pair of bosons $h_a$
can be non small and be approximately equal to each
other [9]. At $\tan\beta\gtrsim 1$, the total width $\Gamma_a$ is small at
any mass $M_a$.

{\it Production of the CP even Higgs boson through a
gauge vertex} was until recently assumed to ensure the
best signal/background ratio and the least inaccuracy in the measurement of its parameters: $W$ fusion at
the LHC, $e^+e^-\to Zh_a$ and $e^+e^-\to \nu\bar{\nu}h_a$ at the ILC,
and $e\gamma\to\nu W^- h_a$, $\gamma\gamma\to W^+H^-h_a$  at the PLC
(photon collider).

In view of Eq. \eqref{SMLsc}, it follows from sum rules (\ref{SR}a)
for all models with any set of their parameters that
experiments on the search for additional Higgs bosons
at the LHC and linear collider in such processes cannot be successful \cite{GinKr13} (such results were
obtained till now only at some sets of parameters in separate
models (see, e.g., recent works  \cite{expanrec}).

Processes of production of the Higgs boson
together with the $t$ quark and through the loop vertex
at the LHC ($gg\to t\bar{t}h_a$, $gg\to h_a$) and at photon collider
($\gamma\gamma \to h_a$, $e\gamma\to eh_a$)
require a separate discussion.

\bu Sum rules (\ref{SR}b)  and relations (\ref{SR})  show that
search for Higgs bosons $h_a$ can be successful in the
$q_1\bar{q}_2\to H^+h_a$, $q\bar{q}\to h_ah_b$ at LHC, $e\gamma\to \nu H^+h_a$, $e^+e^-\to h_ah_b$, $e^+e^-\to H^\pm W^\mp h_a$ at ILC, $\gamma\gamma\to H^\pm W^\mp h_a$ at PLC.

\section{Multidoublet and other models}\label{secnHDM}

By construction, most of the nonminimal models
are symmetric under a certain group of global transformations, which can concern both the scalar and fermion sectors of a model. Spontaneous symmetry
breaking can be responsible for candidates to dark
matter particles, CP violation, etc.

Despite importance of symmetries and numerous
phenomenological studies, it was unknown till now
what symmetry groups can appear in such models,
how these symmetries can be broken, how they are
manifested in the scalar and fermion sectors, and what
their phenomenological consequences are. Only the
simplest variants have been analyzed.

Most of the arising problems in a very wide class of
models were solved in \cite{1}-\cite{11}, where new results were
obtained and new methods for analyzing these models
were developed. These methods are often much more
efficient than traditional approaches and sometimes
reveal some imperfections of these approaches.

A. The problem of classification of symmetry
groups within the scalar sector was studied for theories
with a given set of additional scalar fields. This problem was previously solved only for 2HDM. The Abelian part of this problem was solved for a model with
arbitrary number of doublets \cite{2}. This problem was
completely solved for 3HDM \cite{4,6}. In the process
of solution, a method for analysis of Abelian symmetries was developed on the basis of normal Schmidt
forms, which is appropriate for any models with new
complex fields.

A class of always broken symmetries called frustrated was described \cite{1}. A new geometric method of
the minimization of potentials with such symmetries
was proposed \cite{7}. This method is sometimes better
than more standard methods.

Geometric CP breaking in multidoublet models
was analyzed \cite{10}.

B. Scalar candidates to dark matter particles with
unusual quantum numbers naturally appear in multidoublet models \cite{3}. A new convenient criterion that
cuts off models with metastable vacuum was found for
the two doublet model \cite{8,9}. This criterion can be
verified using LHC data.

C. The problem of classification of symmetries and
study of their consequences was considered in the
quark sector of multidoublet models \cite{11}. The general
results concerning possible symmetries for an arbitrary
number of doublets were obtained. The most interesting examples of 3HDM and 4HDM were analyzed.
The method used leads to results without computer
calculations, opening a direct way to neutrino models.

\section{Inert doublet model}\label{strDM}

The inert doublet model is a serious candidate to
the description of dark matter \cite{inert}. This model is
described by a special variant of 2HDM, where one
Higgs field $\phi_S$ is the same as in the minimal Standard
Model and the other Higgs field $\phi_D$ does not have a
vacuum average and does not interact with fermions.
The corresponding Lagrangian is given by eqs.~\eqref{baslagr},  \eqref{baspot} with $\phi_1\to \phi_S$, $\phi_2\to \phi_D$, $m_{12}=0$, $\ls6=\ls7=0$.

The parameter $\ls 5$ can be taken real and negative. In
addition, we assume $\ls4+\ls 5 < 0$.

Below, the following notation will be used:
\be
 R = \dfrac{\lambda_3+\lambda_4+\lambda_5}{\sqrt{\lambda_1 \lambda_2}},\quad
\mu_1=\dfrac{m_{11}^2}{\sqrt{\lambda_1}},\quad \mu_2=\dfrac{m_{22}^2}{\sqrt{\lambda_2}}.\label{abbr}
\ee
The requirement of the positive potential at large
semiclassical fields imposes the constraints $\ls1, \ls2 > 0$ and
$R > -1$.

The potential has a pair of $Z_2$ symmetries (under
the transformation $\phi_S\to -\phi_S$, $\phi_D \to \phi_D$), which is
called $S$ symmetry and $D$ symmetry (under the transformation $\phi_S\to \phi_S$, $\phi_D \to -\phi_D$). The Yukawa interaction breaks $S$ symmetry, whereas $D$ symmetry is
exact, ensuring conservation of the $D$ parity under this
transformation. The model can pretend to describe
dark matter with the set of parameters such that $\la\phi_D\ra = 0$
and $\la\phi_S\ra = v/\sqrt{2}$. At this minimum of the potential,
$D$ symmetry remains exact and $S$ symmetry is broken
by the Yukawa interaction and by the choice of the
minimum of the potential. The masses of fermions are
expressed in terms of $\la\phi_S\ra$ as in the Standard Model. To
implement this state and a neutral particle for dark
matter, the parameters of the potential should satisfy
the conditions
\be
m_{11}^2>0;\quad\left\{
\begin{array}{lcl}
\mu_1>\mu_2 \;& at& \; R>1,\\
R \mu_1>\mu_2 \;& at& \;
|R|<1.\end{array}\right.\label{inertcond}
\ee
Expansion \eqref{decomp} now becomes
\be
  \phi_S=\begin{pmatrix}G^+\\ \dfrac{v+h+iG^0}{\sqrt{2}}\end{pmatrix},\quad \phi_D=
\begin{pmatrix}D^+\\ \dfrac{D+iD_A}{\sqrt{2}}\end{pmatrix}.\label{decomp1}
\ee
Here, $h$ is the standard Higgs boson with the mass
$M_h= 125$~GeV; $D$, $D_A$, and $D^\pm$ are physical particles
with the masses $M_D$, $M_A$,  $M_\pm\equiv M_+$, respectively;	
\bear{c}
M_{h}^2=\lambda_1v^2,\;\;
M_D^2=
 \dfrac{\sqrt{\lambda_2}(R\mu_1-\mu_2)}{2},\\[2mm]
 M_A^2=M_D^2-v^2\lambda_5,\;\; M_\pm^2=
 M_D^2-v^2\dfrac{\lambda_4+\lambda_5}{2}.
 \eear{masses}
The $P$ parities of $D$ and $D_A$ are opposite to each other.

Particles $D$, $D_A$ and $D^\pm$ are $D$ odd. All other particles are $D$ even. In view of the conservation of the $D$
parity, the lightest of $D$ particles, $D$, can serve as dark
matter.

The masses of $D$ particles are limited by the accelerator and cosmological data \cite{GKKS,krasok}. In particular,
$M_+ > 90$~GeV and $M_A + M_D > 180$~GeV (LEP data).

The scalar particles $D,\, D_A$, and $D^\pm$ interact with
usual particles through covariant derivatives in the
kinetic term $({\cal D}_\mu\phi_D)^\dagger ({\cal D}_\mu\phi_D)$.. Triple interactions with
gauge bosons have the form $D^+D^-\gamma (Z)$, $ D^\pm DW^\mp$, $D^\pm D_AW^\mp$, $DD_AZ$. Interactions with the Higgs
boson $h$ are diagonal in each of the fields $D$, $D_A$ and
$D^\pm$. To describe them, the constant $\ls3$ should be added
to the constants obtained from the measured masses of
the particles. At $M_D < 60$~GeV, measurements of the
invisible decay $h\to DD$, e.g., in the $e^+e^-\to Zh\to ZDD$
reaction will allow determination of $\ls3$.

\bu {\bf Strongly interacting dark matter \cite{14}}. The complete
set of quadratic interactions between $D$ particles has
the form
 \bear{c}
\!\!\!\!\!\dfrac{\ls 2}{8}\cdot\left[ (DD\!+\!D_AD_A)(DD\!+\!D_AD_A\!+\!4D^+D^-)+
 8D^+D^-D^+D^-\right]\,.\eear{DDcoupl}
Differences of the masses of $D$ particles are independent on $\ls2$. Their coupling constants to gauge bosons
and to the standard Higgs boson $h$ are independent on
$\ls2$. Thus, large $\ls2$ values corresponding to strongly
interacting dark matter are not excluded. This interaction at low energies of colliding $D$ particles is repulsion
($\ls2 > 0$). At high energies, attraction can appear with
the formation of resonance states, as was discussed for
possible strong interaction in the Higgs sector of the
Standard Model (see, in particular, recent work \cite{strHiggsnew}).
In this case, further studies are necessary.

\section{Measurement of the masses
of D particles}\label{secnHDM}

The $D^+D^-$ pair is produced at the ILC linear collider with the appropriate energy in $e^+e^-$ collisions
with the cross section close to the cross section for the
$e^+e^- \mu^+\mu^-$ process. These cross sections are very
large for the ILC. Pairs of hadronic jets will be
observed (hadronic decay modes of W), as well as leptons from the $D^+\to DW^+$ decay. The total energy of
the observed particles is much lower than the total
energy of the collision and their total transverse
momentum is quite high. If the energies of hadronic
jets produced from the decay of W could be accurately
measured, the measurement of the boundaries of the
corresponding energy distributions would make it possible to determine the masses of $D^+$ and $D$. Unfortunately, such accurate measurements are impossible.
Only the energies of leptons (e.g., muons) from the
decay of $W$ can be measured well. Simple kinematic analysis shows that the energy distributions of muons have
reliably determined singularities (peaks and kinks),
measurement of which will allow accurate determination of the masses of $D^+$ and $D$ \cite{15}.

This kinematic analysis is independent of the spin
of $D$ particles and can be applied in the case of fermion
dark matter (e.g., if $D$ is neutralino and $D^\pm$ is chargino). In the latter case, the cross section for the
$e^+e^-\to D^+D^-$ process is at least twice as large as the
same cross section for scalar particles. Thus, after the
determination of the masses of particles from the kinematic singularities, even rough measurement of the
cross section will make it possible to determine the
spin of $D$ particles  \cite{15}. More detailed calculations for
the IDM (but without study of aforementioned singularities of the energy spectra of an individual lepton)
were performed in \cite{IDMnewexp}.

\section{Phase states of the Universe
in the IDM}\label{secphstate}

The ground state of the Universe in the period of
cooling after the Big Bang is determined by the minimum of the Gibbs potential
 \be
V_G= Tr\left(V e^{-\hat{H}/T}\right)/Tr\left(
e^{-\hat{H}/T}\right)\,.
 \ee
The Gibbs potential for the IDM in the first nontrivial
approximation at quite high temperatures has the form
of Eq.~\eqref{baspot} with the same coefficients $\ls i$, but with the
mass term depending on the temperature:
\bear{c}
m_{11}^2(T)\!= \! m_{11}^2\!-\!c_1T^2,\;\;
m_{22}^2(T)\!=\!  m_{22}^2\!-\! c_2T^2,\\[3mm]
c_1=\dfrac{3\lambda_1+2\lambda_3+\lambda_4}{12}+\dfrac{3g^2+g^{\prime 2}}{32}+\dfrac{g_t^2+g_b^2}{8},\;\;\;
c_2=\dfrac{3\lambda_2+2\lambda_3+\lambda_4}{12}+\dfrac{3g^2+g^{\prime 2}}{32}.			
\eear{Tempdep}
Here, $g$ and $g'$ are the standard gauge coupling constants of the electroweak theory while $g_t\approx 1$ and $g_b\approx
0.025$ are the coupling constants of the $t$ and $b$ quarks
with the Higgs boson\fn{This is the high-temperature approximation. At moderate energies, dependencies become more complicated. We hope that our
calculations correctly describe the general qualitative picture of
phenomena.}. In view of positivity constraints, $c_1 + c_2 > 0$ \cite{GKKS}.

Conditions \eqref{inertcond} can be violated at finite temperatures. In this case, the properties of the ground state
can be significantly different from the present properties  \cite{GIK09}.

All extrema of the IDM potential are easily
obtained because it has a simple structure. These
extrema are listed below with the corresponding values
$v_i=\sqrt{2}\la\phi_i\ra$   and energy of the ground state ${\cal E}$ \cite{GKKS}:
\bes\label{eqvacua}\bea{\pmb {EWs}}:& v_D=0,\quad v_S=0,\quad\quad {\cal E}_{EWs}=0;&\label{Sol0bas}\\
{\pmb {I_1}}:& v_D=0,\; v_S^2=\dfrac{\mu_1}{\sqrt{\lambda_1}},\quad
     {\cal E}_{I_1}=-\dfrac{\mu_1^2}{8};&
     \label{solAbas}\\
{\pmb {I_2}}:& v_S=0,\; v_D^2=\dfrac{\mu_2}{\sqrt{\lambda_2}},\quad
     {\cal E}_{I_2}=-\dfrac{\mu_2^2}{8};&
          \label{solBbas}
     \eea
\be
\!\!\!{\pmb M}:\left\{
    \begin{array}{c}
v_S^2\!=\!\dfrac{\mu_1-R\mu_2}{\sqrt{\ls 1}(1-R^2)},\;\;
v_D^2\!=\!\dfrac{\mu_2-R\mu_1}{\sqrt{\ls 2}(1-R^2)},\;\;
v_M^2= v_S^2+v_D^2\,,\\[3mm]{\cal E}_{M}=-\dfrac{\mu_1^2+\mu_2^2-2R\,\mu_1\mu_2}{8(1-R^2)}.
    \end{array}\right.
\label{Nextr1}\ee
\ees
(If one of the quantities $v_S^2$ and $v_D^2$ in \eqref{Nextr1}
is negative, the extremum $\pmb M$ does not realized.)

The ground state (vacuum) corresponds to the
extremum with the lowest energy $\cal E$. All possible vacuum states \eqref{eqvacua} are briefly described below.

\bu {\bf EW symmetric state $\pmb{ EWs}$.}
This extremum exists at
any parameters of the potential. It conserves the $D$ and $S$
symmetries of the potential, and is a minimum realizing vacuum at
\begin{equation}
 m_{11}^2<0,\qquad m_{22}^2<0.
  \end{equation}
In this vacuum all particles are massless except for
scalar doublets with the masses $|m_{11}|/\sqrt{2}$ and $|m_{22}|/\sqrt{2}$.

\bu {\bf Inert state $\pmb{I_1}$}. The properties of this state are
described in Section~\ref{strDM}. It is the ground state (vacuum)
under conditions  \eqref{inertcond}.

In two other phases described
below, there are no particles having the properties of
dark matter.

\bu {\bf Inert-like state $\pmb{I_2}$}. At first glance, the properties of
this state are similar to the properties of the inert state
with the change $D \leftrightarrow S$. However, $S$ particles in this
state that are similar to $D$ particles in the inert state
interact with fermions, which remain massless. The
conditions of realization of the state I2 as vacuum are
similar to conditions  \eqref{inertcond}:
\be
m_{22}^2>0;\quad\left\{
\begin{array}{l}
\mu_2>\mu_1 \; at \; R>1,\\
R \mu_2>\mu_1 \; at \;
|R|<1.\end{array}\right.\label{inertlikecond}
\ee

\bu {\bf Mixed state M}. In this state, both $D$ and $S$ symmetries are broken. Here, the masses of fermions are
expressed via their present values as $m_{f,M}=m_f(v_S/v)$. The standard expansion near this extremum
gives Goldstone bosons $G^\pm$, $G^0$, charged Higgs
bosons $H^\pm$, a pseudoscalar particle $A$ with the masses
\be
 M_{H^\pm}^2=-\dfrac{\lambda_4+\lambda_5}{2}v_M^2\,,\quad
 M_A^2=-v_M^2\lambda_5, \label{masses1}
 \ee
and two scalar particles $h$ and $H$ with the masses
\bear{c}
\!M_{h,H}^2\!=\!\dfrac{\lambda_1v_S^2\!+\!\lambda_2v_D^2\!\pm\sqrt{\Delta}}{2},\quad
\Delta=
(\lambda_1v_S^2\!+\!\lambda_2v_D^2)^2\!-\!4\ls 1\ls 2(1-R^2)v_S^2v_D^2.\eear{massesC}

According to Eq.~\eqref{eqvacua}, if the mixed state gives a
minimum of the potential, this minimum is global;
i.e., it is vacuum. This occurs under the conditions
$|R| < 1$ and
\bear{lc}
\!\!{\rm at}\;\;\;\; 1>R>0: & 0<R\mu_1<\mu_2<\dfrac{\mu_1}{R} ;\\[4mm]
\!\!{\rm at}\;\; 0>R>-1:&
\mu_2>R\mu_1,\;\;\mu_2>\dfrac{\mu_1}{R}.
\eear{Ccond2}

\bu {\bf Degeneracy of the mixed state} \cite{14}. The initial
Lagrangian is symmetric under replacement $\phi_D\to -\phi_D$.
Hence, extremum \eqref{Nextr1} is degenerate in the sign of
$\la\phi_D\ra\equiv  v_D$ and there are states $MP+$ with
$\la\phi_D\ra=  |v_D|$
 and
$M_-$ with $\la\phi_D\ra=  -|v_D|$. In these states, the signs of the
coupling constants of the heavy Higgs boson $H$ with
the gauge boson $Z$ are different. This difference can be
detected in the $t\bar{t}\to WW$ process, where the $Z$ and $H$
exchanges interfere.

The height of the energy barrier between the states $M_+$ and $M_-$ is
given by the energy of a saddle point between them,
i.e., by the extremum next in magnitude in
Eqs.~\eqref{eqvacua}. The height of this barrier ${\cal E}_B$ is
\be
{\cal E}_B=\left\{\begin{array} {lc}
\!\!\!{\cal E}_{I1}-{\cal E}_M\equiv\dfrac{(\mu_2-R\mu_1)^2}{8(1-R^2)} &at\; \mu_1>\mu_2,\\
\!\!\!{\cal E}_{I2}-{\cal E}_M\equiv \dfrac{(\mu_1-R\mu_2)^2}{8(1-R^2)} &at\; \mu_2>\mu_1.
 \end{array}\right.\label{EB}
 \ee

\section{Evolution of the phase states
of the Universe}

The possible phase history of the Universe within
the IDM will be described below under the assumption that the present state is the inert vacuum I1.

It is convenient to use the phase plane $\left(\mu_1(T),\,\mu_2(T)\right)$, where
\be
\mu_1(T)= m_{11}^2(T)/\sqrt{\lambda_1},\quad \mu_2(T)= m_{22}^2(T)/\sqrt{\lambda_2}\,, \label{plane}
\ee
with the functions $m_{ii}^2(T)$  defined by Eq.~\eqref{Tempdep} (the
argument $T$ is omitted in the present $\mu_i$ values). The
present state of the Universe is shown in the figures by
the black point  $Pn=(\mu_1,\,\mu_2)$, where $n$ is the number of
the point. Now, in the inert phase, $\mu_1 > 0$. The
parameter $\mu_2$ can be both positive (points $P1$ and $P3$)
and negative (points $P2$ and $P4$). At $R > 0$, we have
$c_2 > 0$ and $c_1 > 0$  \cite{GKKS}, and  in the
initial state of the Universe ($T\to\infty$) we have $m_{11}^2<0$ and $m_{22}^2<0$; i.e., this initial state has electroweak symmetry.

In accordance with  Eq.~\eqref{Tempdep}, the evolution of the Universe is described by the ray nm, which ends at the
point $Pn$ (m is the number of the type of ray). The
arrow in this ray corresponds to an increase in the time
(a decrease in the temperature). The directions of the
rays are given by the parameters $\tilde{c}_1\equiv c_1/\sqrt{\lambda_1}$ and $\tilde{c}_2\equiv c_2/\sqrt{\lambda_2}$.

\bu {\bf For the case $\pmb{R > 1}$}, the phase diagram is shown in
left plot of Fig. 1. It contains one quadrant with the $EWs$ phase
and two sectors with the phases $I_1$ and $I_2$. These sectors
are separated by {\it the phase transition line} $\mu_1(T) = \mu_2(T)$ (thin black line). Two typical possible present day states are
represented by the points $P1$ ($\mu_2 > 0$) and $P2$ ($\mu_2 < 0$) while the
possible types of evolution are given by the rays 11, 12,
and 21.

$\Box$ \textbf{Rays 11 ($\pmb{c_2/c_1>m_{22}^2/m_{11}^2}$) and 21 ($\pmb{m_{22}^2<0}$)}. Evolution starts from the EW symmetric phase $EWs$. As in
the Standard Model, the Universe occurs in the
present inert phase after the single EWSB second
order phase transition at  $m_{11}^2(T)=0$, i.e., at the temperature
\be
 T_{EWs,1}=\sqrt{m_{11}^2/c_1}\,,\label{TEWSBI_1}
\ee
with the order parameter $\eta_{EW1}\propto\la\phi_S\ra=v_S$, which is
represented by the mass of the usual Higgs boson $M_h$.

$\Box$ \textbf{Ray 12 ($\pmb{c_2/c_1<m_{22}^2/m_{11}^2}$)}. Evolution starts from
the $EWs$ phase. Then, the Universe goes to the inert-like phase $I_2$   at
$m_{22}^2(T)=0$, i.e., at the temperature
\be
 T_{EWs,2}=\sqrt{m_{22}^2/c_2}\,,\label{TEWSBI_2}
\ee
That is EWSB phase transition of
the second order
with the order parameter $\eta_{EW2}\propto\la\phi_D\ra=v_D$, which is
represented by the mass of the Higgs boson $M_{hD}$.

Upon the further cooling,  the Universe goes into
the inert phase $I_1$ at $\mu_2(T) = \mu_1(T)$, i.e., at the
temperature
\be
T_{2,1}=\sqrt{ \dfrac{\mu_1-\mu_2}{\tilde c_1 -\tilde c_2}} \,. \label{TI1I2}
\ee
That is a first order phase transition with the latent heat
\bear{c}
 Q_{I_2\to I_1}=\left. T\dfrac{\pa{\cal E}_{I_2}}{\pa T}-T\dfrac{\pa{\cal E}_{I_1}}{\pa T}
\right|_{\mu_2(T)\to\mu_1(T)}
=({m_{22}^2 c_1-m_{11}^2 c_2})T^2_{2,1}/(4\sqrt{\ls 1 \ls 2})\,.
\eear{heat}

Near this phase transition, the usual formation of
bubbles of a new phase occurs, which is manifested in
the present spatial structure of the cosmic microwave
background. This transition can occur at much lower
temperature than the EWSB phase transition in the
Standard Model.

\begin{figure}[h!]
\begin{center}
  \includegraphics[width=0.45\textwidth]{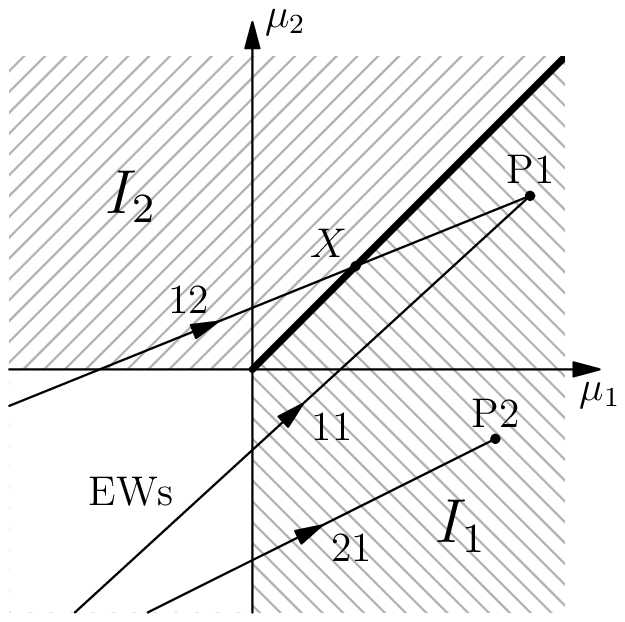}\hspace{3mm}
\includegraphics[width=0.45\textwidth]{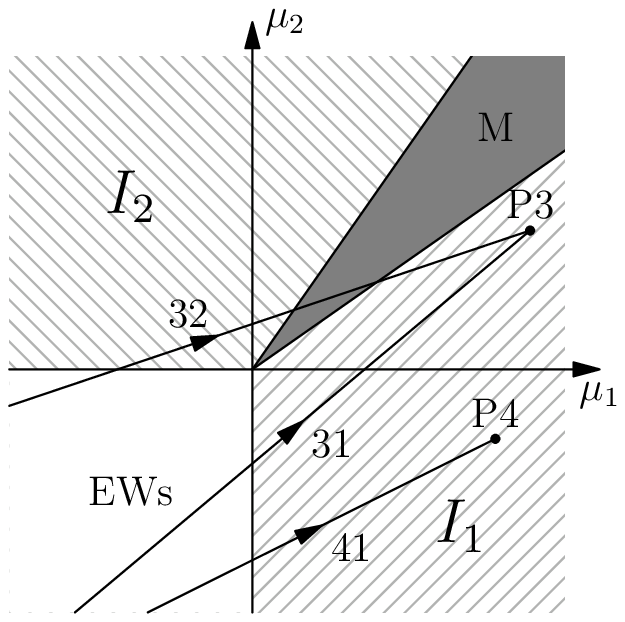}
    \caption{\it  Possible paths of the phase evolution of the Universe. Left -- at $R > 1$, right -- at $1 >R > 0$}
\end{center}
\end{figure}

\bu {\bf For the case $\pmb{1>R > 0}$}, the phase diagram is shown in
right plot of Fig. 1. In addition to the phases of the preceding
case (left plot), a new (grey in figure) sector with the mixed M phase
appears in the upper right quadrant; according to
Eq.~\eqref{Ccond2}, its boundaries are given by the relations
\be
0< R\mu_1(T)<\mu_2(T)< \mu_1(T) /R.  \label{Mphaseb}
\ee
Two typical possible present day states are shown by the
points $P3$ ($\mu_2 > 0$) and $P4$ ($\mu_2 < 0$) and the rays 31, 32,
and 41 represent the possible paths of evolution.

$\Box$ Phase evolution for the 31 and 41 rays is similar to that
at  rays 11 and 21 in left plot.

$\Box$ {\bf Ray 32 ($\pmb{c_2/c_1<m_{22}^2/m_{11}^2}$)}. The Universe starts from the $EWs$ state. the Universe goes to the inert-like phase $I_2$   at
$m_{22}^2(T)=0$, i.e., at the temperature $ T_{EWs,2}$, given by eq.~\eqref{TEWSBI_2}. Just as in the previous case that is EWSB phase transition of the second order
with the order parameter $\eta_{EW2}\propto\la\phi_D\ra=v_D$, which is represented by the mass of the Higgs boson $M_{hD}$. With further cooling, the
Universe passes through the mixed phase $M$ to the
present day inert phase $I_1$.

The phase transition $I_2\to M_\pm$ occurs at the temperature
\be
T_{2,M}= \sqrt{({ \mu_1-R\mu_2})/({ \tilde c_1 - R\tilde c_2})}\,. \label{T2M}
\ee
That is the second order phase transition with the
order parameter $\eta_{I2M}\propto\la\phi_D\ra=v_D$, the latter is given by
the mass of the "usual" Higgs boson $M_{hD}$ at in the inert-like phase $I_2$ and by the mass of the lightest Higgs boson $M_h$
in the mixed phase $M_\pm$.
The energy barrier between the
phases $M_+$ and $M_-$, given by Eq.~ \eqref{EB} increases as $\eta_{I2M}^4$
on departure from the transition point.

When the temperature decreases below $T_{2,1}$ \eqref{TI1I2}, we turn to the region $\mu_1 > \mu_2$  with the change
of the order parameter from $\eta_{I2M}$ to $\eta_{MI1}\propto\la\phi_S\ra= v_S$,
which is represented by the mass of the modern Higgs
boson $M_h$ in the inert phase $I_1$ and by the mass of the
lightest Higgs boson $M_h$ in the mixed phase $M_\pm$. After
that, the energy barrier decreases as $ \eta_{MI1}^4$. At the
temperature
\be
T_{M,1}= \sqrt{({R\mu_1- \mu_2})/({R\tilde c_1 -  \tilde c_2})}\label{TM1}
\ee
the $M\to I_1$ phase transition occurs. That is the second order transition
with the order parameter$\eta_{MI1}$.

When approaching the transition temperatures
$T_{phtr}=T_{2,M}$  and $T_{M,1}$, the masses of bosons representing the order parameters tend to zero as $M_a^2=A_a|T^2-T^2_{phtr}|$
with different coefficients $A_a$.

Near these transitions, large critical fluctuations
appear; their footprints can be sought in the present
spatial structure of the cosmic microwave background.
In particular, the mixed phase $M$ near $I_2\to M$ transitions is built from domains of 3 types, that are  $I_2$ phase domains
(obliged by fluctuations of the
temperature and density), and domains $M_+$ and $M_-$
with the height of walls between domains $\propto \eta_{2,M}^4$.
The spatial distribution of these domains varies continuously. The characteristic correlation radius of a
domain is $R_c(T)\propto 1/\eta_{2,M}\propto 1/\sqrt{|T^2-T_{2,M}^2|}$.

With further cooling, $I_2$ domains become less
energy preferred, their number is decreased, the $M_+$ and $M_-$ domains are "tempered"; i.e., the height of walls between them increases.
Domains are transformed to bubbles with the surface
tension $\sigma_s\sim E_bR_c$. The curved surface of such bubbles
produces pressure $\sim \sigma_s/r$, where $r$ is the local radius of
curvature. Large domains absorb small domains owing
to this pressure. The local velocity of a domain wall is
about the speed of light $c$. At the same time, the global
mixing process is a slow diffusion process with the
characteristic time $\sim (R/c)\sqrt{R/R_c}$, where $R$ is the
characteristic dimension of the inhomogeneity of the
Universe.

When the temperature decreases below $T_{2, 1}$, evolution of domains proceeds by the inverse way with
the change in the order parameter $\eta_{2,M}\to \eta{M,1}$.

\bu {\bf At $\pmb{0 > R > -1}$}, the phase diagram is similar to that in right plot of Fig. 1,
but with an important change: {\it the  region of
the mixed phase covers the entire upper right quadrant, expands beyond it}, and is located between the
rays $\mu_2>\mu_1/R$ and $\mu_2>\mu_1R$. Furthermore, the inequality $c_2/c_1 < 0$ can be satisfied in this region. In this case  electroweak
symmetry of the initial state of the Universe is broken
and new types of phase evolution appear starting from the initial inert-like phase $I_2$ and arriving at the
present inert phase $I_1$ either through the mixed phase
(as the 12 ray) or through the intermediate phase with
electroweak symmetry (two second order phase transitions).\\

\bu The presented picture leaves many interesting
questions for further studies.

{\it Variants with the transition through the mixed phase
or with a first order phase transition.}

 Depending on the
parameters of the model, the last phase transition in
the Universe can occur at quite low temperature.

What
is the relation between the rate of diffusion equalization of fluctuations and the expansion rate of the Universe? Are there scales at which inhomogeneities of the
Universe exist at present?

{\it Variant with the transition through the mixed phase.}

How long do footprints of "tempering" of domains hold
in the inert phase?\\

The reviewed studies were supported by the Russian Foundation for Basic Research (project  11-02-00242 and preceding projects), by the Council of
the President of the Russian Federation for Support of
Leading Scientific Schools
project  NSh-3802.2012.2), and by the Division of
Physical Sciences, Russian Academy of Sciences
(program "Study of the Higgs Boson and Exotic Particles at the LHC").

\end{document}